\documentclass[aps,preprintnumbers,showpacs,twocolumn,superscriptaddress,floatfix,nofootinbib]{revtex4-1}

\usepackage{latexsym} 
\usepackage{amssymb} 
\usepackage{amsfonts}
\usepackage{amsmath} 
\usepackage{bm} 
\usepackage{mathtools}
\usepackage{envmath}
\usepackage{color}
\usepackage{times} 
\usepackage{hyperref}
\usepackage{float}
\usepackage{latexsym}
\usepackage{envmath}

\def\prl{Phys. Rev. Lett.}
\def\prd{Phys. Rev. D}
\def\cqg{Class. Quantum Grav.}

\def\omrot{\Omega}

\usepackage{amssymb,amsmath}
\usepackage{graphicx} 

\begin{document}
\title{Critical collapse of rotating radiation fluids}

\author{Thomas W. \surname{Baumgarte}}\affiliation{Department of Physics and Astronomy, Bowdoin College, Brunswick, ME 04011, USA}

\author{Carsten \surname{Gundlach}}\affiliation{Mathematical Sciences, University of Southampton, Southampton SO17 1BJ, United Kingdom}


\begin{abstract}
We present results from the first fully relativistic simulations of the critical collapse of rotating radiation fluids.  We observe critical scaling both in subcritical evolutions, in which case the fluid disperses to infinity and leaves behind flat space, and in supercritical evolutions that lead to the formation of black holes.  We measure the mass and angular momentum of these black holes, and find that both show critical scaling with critical exponents that are consistent with perturbative results.  The critical exponents are universal; they are not affected by angular momentum, and are independent of the direction in which the critical curve, which separates subcritical from supercritical evolutions in our two-dimensional parameter space, is crossed.    In particular, these findings suggest that the angular momentum decreases more rapidly than the square of the mass, so that, as criticality is approached, the collapse leads to the formation of a non-spinning black hole.   We also demonstrate excellent agreement of our numerical data with new closed-form extensions of power-law scalings that describe the mass and angular momentum of rotating black holes formed close to criticality. 
\end{abstract}


\maketitle

Critical phenomena in gravitational collapse, first reported in the seminal work of Choptuik \cite{Cho93}, refer to properties of solutions to Einstein's equations close to the threshold of black-hole formation (see \cite{Gun03,GunM07} for reviews).   Consider a family of initial data, for a given matter model, parameterized by a parameter $p$.  {\em Supercritical} data will evolve to form a black hole, while {\em subcritical} data will not.  The onset of black-hole formation occurs at some critical value of the parameter, say $p_*$.  The mass of black holes formed by supercritical data then scales with
\begin{equation} \label{mass_scaling}
M \simeq C_{ M} \, | p - p_* |^{\gamma_{ M}},
\end{equation}
where the critical exponent $\gamma_{ M}$ depends on the matter model, but not on the specifics of the initial data or their parametrization (depending on the choice of the parameterization, super\-critical solutions may correspond to $p > p_*$ or $p < p_*$).    For subcritical data, for which the fluid disperses to infinity and leaves behind flat space, the maximum value of the spacetime curvature attained during the evolution also follows critical scaling (see \cite{GarD98}).  For perfect fluids, for example, Einstein's equations relate this maximum curvature to the maximum value of the density $\rho$ encountered during the evolution, leading to a scaling
\begin{equation} \label{rho_scaling}
\rho_{\max} \simeq C_{\rho} \, | p - p_* |^{- 2 \gamma_{\rho}}
\end{equation}
for subcritical data.  On dimensional grounds, we must have $\gamma_\rho = \gamma_M$.  Moreover, in the strong-field region prior to black-hole formation, the solution approaches a self-similar critical solution which also depends on the matter model but not the initial data.

Choptuik's discovery of these critical phenomena launched an entire new field of research.   Soon after his announcement, which was based on simulations of massless scalar fields, similar phenomena were reported in the collapse of vacuum gravitational waves \cite{AbrE93} and radiation fluids \cite{EvaC94}, followed by numerous other numerical, analytical and perturbative studies for different matter models, asymptotics, and number of spacetime dimensions (we again refer to \cite{Gun03,GunM07} for reviews).   In particular, these studies revealed that for some matter models, including scalar fields, the self-similarity of the critical solution is discrete, while for others, including perfect fluids, it is continuous.  For many models
the critical exponent $\gamma_{ M}$ can also be found semi-analytically in perturbation theory (e.g.~\cite{KoiHA95,Mai96}).

The vast majority of these studies, however, was performed under the assumption of spherical symmetry.  In particular, despite the tremendous recent progress in numerical relativity, only few numerical simulations of aspherical critical collapse have been performed (e.g.~\cite{AbrE93,ChoHLP03b,ChoHLP04,Sor11,HilBWDBMM13,HeaL14,BauM15}).   This is even more surprising as several interesting questions cannot be addressed in spherical symmetry.  One such question concerns the angular momentum in the collapse of rotating matter.   To date, the only fully nonlinear and relativistic study of the role of angular momentum in critical collapse was performed by Choptuik {\it et.al.} \cite{ChoHLP04}, who considered a complex scalar field and constructed initial data carrying angular momentum in such a way that the resulting stress-energy tensor was axisymmetric.  This approach leads to an aspherical density distribution, so that it did not allow for an exploration of the angular momentum's role in perturbing spherical critical collapse.  

Gundlach \cite{Gun02} (see also \cite{Gun98b}) considered nonspherical perturbations of the critical solution for perfect fluids, whose pressure $P$ is related to the density $\rho$ by an equation of state $P = \kappa \rho$.  These studies showed that, close to the onset of black-hole formation, the angular momentum scales with
\begin{equation} \label{ang_mom_scaling}
J \simeq C_{ J} \, | p - p_* |^{\gamma_{ J}},
\end{equation}
where the critical exponent $\gamma_{ J}$ is related to $\gamma_{ M}$ by
\begin{equation} \label{gammas}
\gamma_{ J} = \frac{5 \, (1 + 3 \kappa)}{3 \, (1 + \kappa)} \,\gamma_{ M}
\end{equation}
for $1/9 < \kappa \lesssim 0.49$ (see eq.~(23) in \cite{Gun02}).  For a radiation fluid with $\kappa = 1/3$ we obtain $\gamma_{ J} = 2.5 \, \gamma_{ M}$, or $\gamma_{ J} \simeq 0.8895$ for the analytical value $\gamma_{ M} \simeq 0.3558$ \cite{KoiHA95,Mai96}.  Combining the scaling relations (\ref{mass_scaling}) and (\ref{ang_mom_scaling}) we also have 
\begin{equation} \label{J_versus_M}
J \propto M^{\gamma_{ J}/\gamma_{ M}},
\end{equation}
which shows that the angular momentum should decrease more rapidly than the square of the mass as criticality is approached, so that, in this limit, the forming black hole should be non-spinning.  This behavior is similar to the observation that charge does not affect the critical solution in the critical collapse of charged scalar fields \cite{GunM96,HodP97}.  

In this paper we report on what we believe are the first fully relativistic simulations of the gravitational collapse of rotating radiation fluids (but see \cite{Agu15} for a study in Newtonian gravity).  We confirm the above relations for the critical exponents, to within the accuracy of our simulations, and demonstrate excellent agreement with new closed-form extensions of power-law scalings that describe the mass and angular momentum of rotating black holes formed close to criticality \cite{GunB16}.

We consider a radiation fluid with $P = \rho / 3$, i.e.~$\kappa = 1/3$, and generalize the initial data adopted in \cite{EvaC94} by allowing the fluid to carry angular momentum (see eqs.~(\ref{rho_init}) and (\ref{S_init}) below).   We then evolve these data with the Baumgarte-Shapiro-Shibata-Nakamura (BSSN) formulation \cite{NakOK87,ShiN95,BauS98}, which adopts both a 3+1 decomposition as well as a conformal rescaling $\gamma_{ij} = \psi^4 \bar \gamma_{ij}$ of the spatial metric $\gamma_{ab} \equiv g_{ab} + n_a n_b$.  Here $g_{ab}$ is the spacetime metric, $n^a$ the normal vector on spatial slices, $\psi$ the conformal factor, and $\bar \gamma_{ij}$ the conformally related metric.  The extrinsic curvature $K_{ij}$ is related to the time derivative of the spatial metric.  We solve the resulting equations in spherical polar coordinates \cite{BauMCM13,MonBM14,BauMM15}, imposing 1+log slicing and a Gamma-driver condition.  The code makes no symmetry assumptions, but we run it here assuming both axisymmetry and a symmetry across the equatorial plane.  In \cite{BauM15} we used this code to study critical phenomena in the aspherical collapse of a radiation fluid; those calculations also serve as a calibration of our code for the calculations presented here.   We use a logarithmic grid in the radial direction (see App.~A in \cite{BauM15}), and also allow for a radial regridding to zoom in on the critical solution.  In most simulations our radial resolution at the origin is $\Delta r \simeq 5 \times 10^{-3}$ initially, but $\Delta r \simeq 5 \times 10^{-4}$ at late times.   As in \cite{BauM15} we have found it sufficient to use only $N_\theta = 12$ angular gridpoints to resolve one hemisphere.

Following \cite{EvaC94} we choose maximally sliced (i.e.~$K \equiv \gamma^{ij} K_{ij} =0$) and conformally flat (i.e.~$\bar \gamma_{ij} = \eta_{ij}$) initial data with an initial density distribution
\begin{equation} \label{rho_init}
\rho_n \equiv n_a n_b T^{ab}  = \frac{\eta}{2 \pi^{3/2} R_0^2} \exp{\left(- (\psi^2 r/R_0)^2 \right)},
\end{equation}
where $T^{ab}$ is the stress-energy tensor.  Here $\rho_n$ is the density as observed by a normal observer; the density as observed by an observer comoving with the fluid is $\rho \equiv u_a u_b T^{ab}$, where $u^a$ is the fluid's four-velocity.  For spherically symmetric data we may interpret $R \equiv \psi^2 r$ as the areal radius.  In \cite{BauM15} we considered aspherical deformations of this density distribution; here we instead consider rotating fluids with an initial momentum density
\begin{equation} \label{S_init}
S^\varphi \equiv - \gamma^{\varphi j} n^i T_{ij} = \frac{4}{3} \rho_n \frac{\omrot}{1 + (\psi^2 r/R_0)^2}
\end{equation}
and $S^r = S^\theta = 0$.  Given $S^\varphi$ we solve the momentum constraints for a vector potential $W^\varphi$ from which the trace-free part of the extrinsic curvature $A_{ij}$ can be computed (see, e.g., Box 3.1 in \cite{BauS10}.)  Solving the Hamiltonian constraint then yields the conformal factor $\psi$.  We solve the coupled set of equations iteratively, updating the sources (\ref{rho_init}) and (\ref{S_init}) between iterations, until the solution has converged to a desired accuracy.   For a radiation fluid, the above fluid variables are identical to the corresponding conserved fluid variables used in our hydrodynamical scheme, from which the primitive fluid variables $\rho$, $P$ and $v^\varphi$ can then be recovered.   For $\omrot = 0$ we also have $\rho = \rho_n$ initially, so that the above data reduce to the initial data of \cite{EvaC94} in that limit.  To complete the initial data we choose a ``precollapsed" lapse $\alpha = \psi^{-2}$ and zero shift at the initial time.
 
In (\ref{rho_init}) and (\ref{S_init}), $\eta$ parametrizes the overall amplitude of the density, and $\omrot$ the rotation rate.  
A third parameter, $R_0$, determines the length-scale of the problem.  We fix our code units by setting $R_0 = 1$; all dimensional quantities are hence expressed in units of $R_0$.   

\begin{figure}[t]
\includegraphics[width=3.5in]{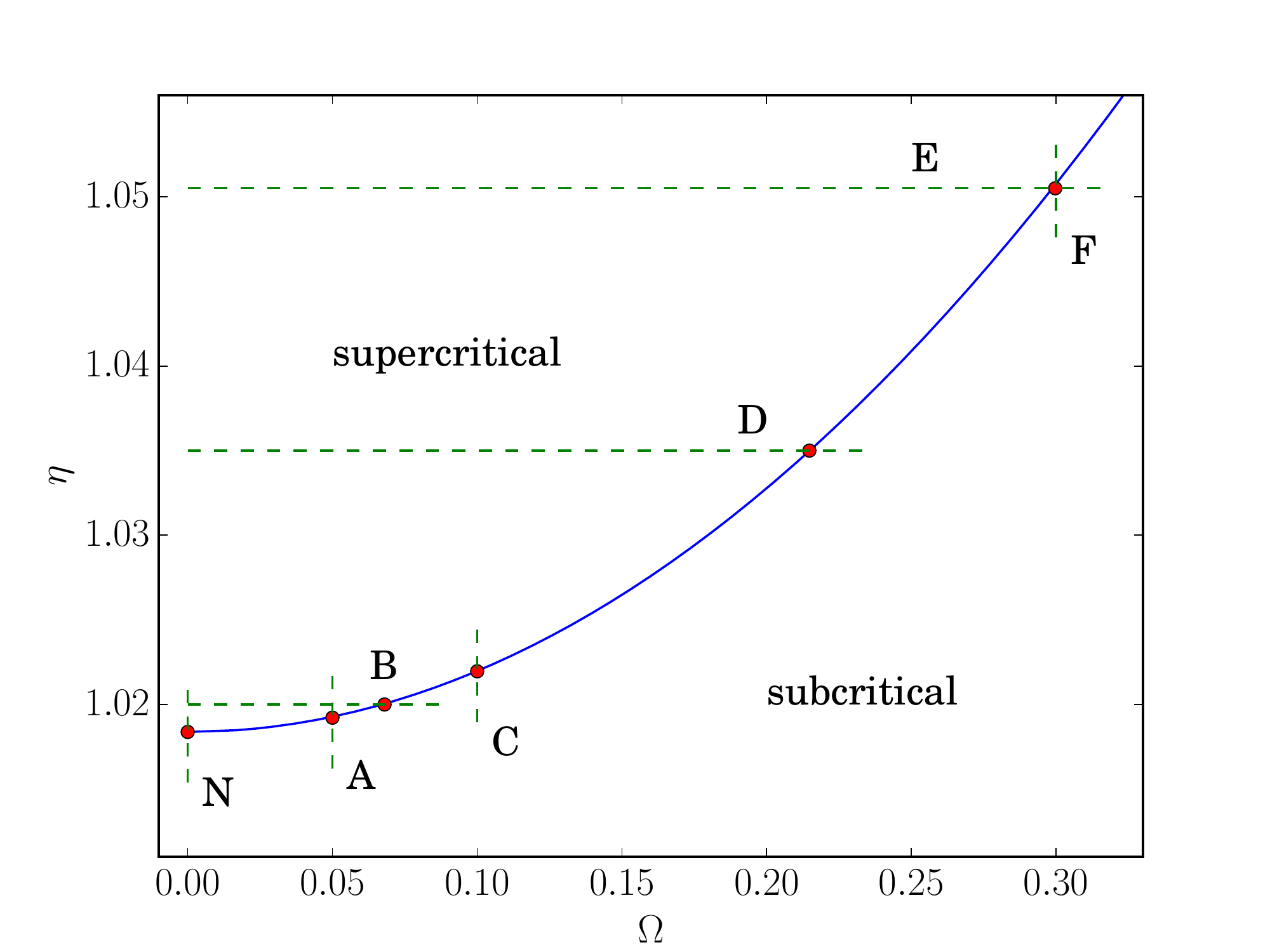}
\caption{A sketch of our numerical sequences (see also Table \ref{Table1} for parameters and results).  The green dashed lines show our sequences, with the red dots marking the critical points.  The blue line is the fit (\ref{eta_crit_fit}) through these points and represents the critical curve that separates supercritical from subcritical configurations.  Supercritical data have $\eta > \eta_*$ for sequences of constant $\omrot$, but $\omrot < \omrot_*$ for sequences of constant $\eta$.} 
\label{Fig:sketch} 
\end{figure}

\begin{figure}[t]
\includegraphics[width=3.5in]{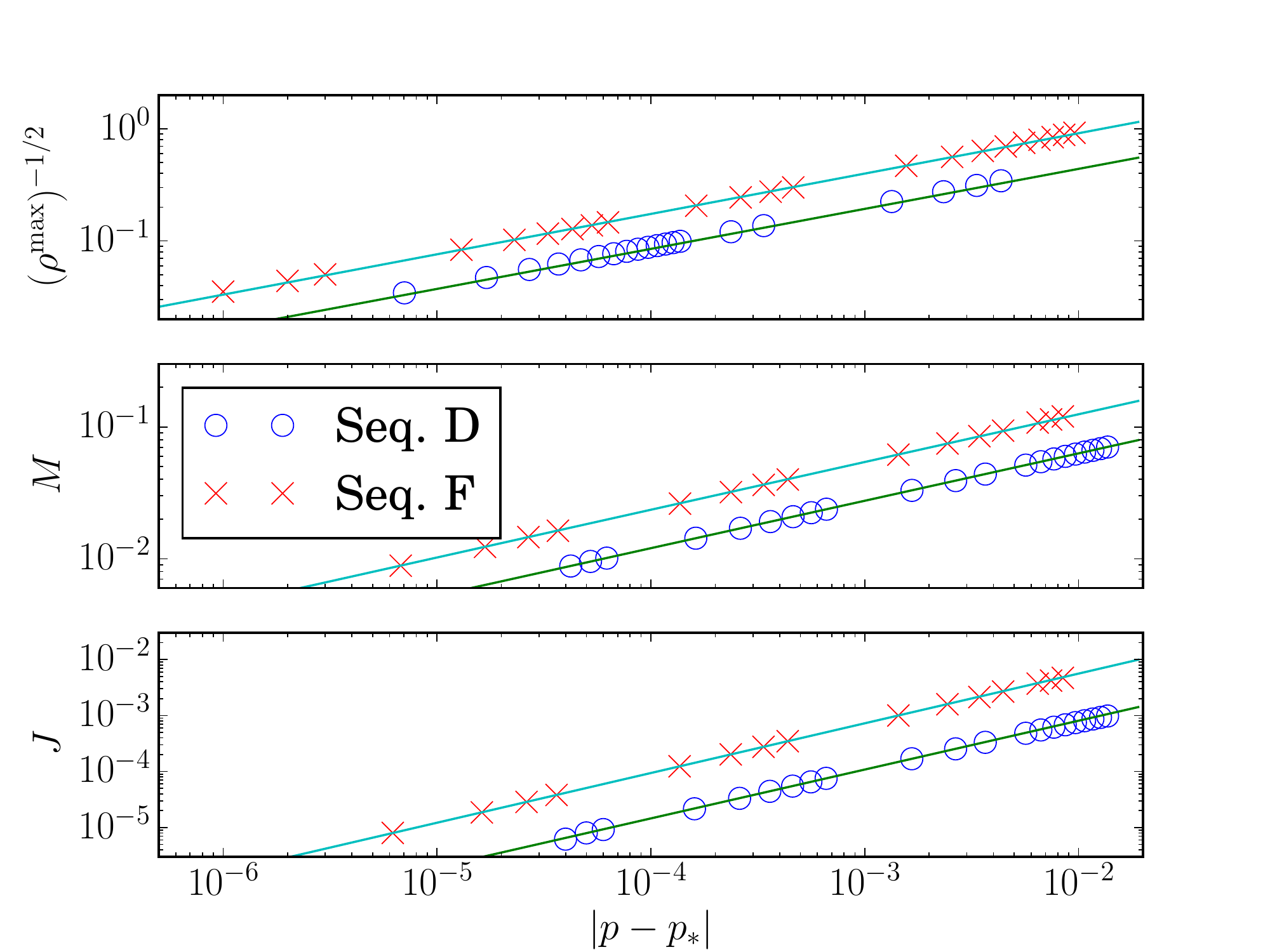}
\caption{Scalings for sequences $D$ and $F$.  The upper panel shows results for the maximum density $\rho_{\rm max}$ encountered in subcritical evolutions, while the lower two panels show results for the mass $M$ and angular momentum $J$ of black holes formed in supercritical evolutions.  The crosses and dots denote our numerical results; the lines represent fits of the scaling laws (\ref{mass_scaling}), (\ref{ang_mom_scaling}) and (\ref{rho_scaling}) based on the parameters listed in Table \ref{Table1}.  The parameter $p$ corresponds to $\omrot$ in sequence $D$, but to $\eta$ in sequence $F$.}
\label{Fig:Fig2} 
\end{figure}

In the following we explore several different sequences through our two-dimensional parameter space, as indicated by the dashed green lines in Fig.~\ref{Fig:sketch}.   Six of these sequences (labeled $A$ - $F$) are for rotating configurations; a seventh (labeled $N$) is the non-rotating limit \cite{EvaC94} (also \cite{BauM15}).

We first explore the threshold of black-hold formation by exploring sequences $A$ - $F$ in the vicinity of criticality.  For each sequence we vary the parameter that is not being held constant to bracket the critical value of this parameter (the red dots in Fig.~\ref{Fig:sketch}).   We summarize the parameters and results for these sequences in Table \ref{Table1}.    As one might expect, rotation provides centrifugal support to the fluid approximately proportional to the square of the rotation rate, so that with $\omrot \ne 0$ a black hole forms only 
for larger values of $\eta$ than with $\omrot = 0$.   This is borne out by the blue line in Fig.~\ref{Fig:sketch}, which represents the leading-order fit
\begin{equation} \label{eta_crit_fit}
\eta_*(\omrot_*) = \eta_{*0} + 0.36 \, \omrot_*^2
\end{equation}
through the critical points $(\eta_*,\omrot_*)$ and marks the critical curve that separates the supercritical from subcritical parts of our parameter space.  Here $\eta_{*0} \simeq 1.0184$ denotes the critical value for zero rotation. 

For supercritical data, which correspond to $\eta > \eta_*$ for sequences of constant $\omrot$, but to $\omrot < \omrot_*$ for sequences of constant $\eta$, we find an apparent horizon \cite{ShiU00b} and measure its irreducible mass $M_{\rm irr}$ and angular momentum $J$  (see \cite{DreKSS03}) once their values have settled down to approximately constant values. (Angular momentum that does not end up in a black hole is carried away by the dispersing fluid.)  Assuming that the new black hole is a Kerr black hole we then compute the Kerr mass $M = M_{\rm irr} (1 + (J/M_{\rm irr}^2)^2/4)^{1/2}$.   Fitting our numerical data for the mass $M$ and angular momentum $J$ to the scaling relations (\ref{mass_scaling}) and (\ref{ang_mom_scaling}) then yields the critical exponents $\gamma_{M}$, $\gamma_{J}$.  For subcritical data we fit the maximum encountered densities $\rho_{\rm max}$ to the scaling relation (\ref{rho_scaling}) to find $\gamma_{\rho}$.  We show examples of these scalings for sequences $D$ and $F$ in Fig.~\ref{Fig:Fig2}, and we list all our results in Table~\ref{Table1}.

\begin{table*}[t]
\begin{tabular}{c|l|l|l|l|l|l|l|l}
	& fixed par. & crit.\ value & $M_{\rm tot}$ & $J_{\rm tot}$ & $\gamma_{\rho}$ & $\gamma_{M}$ & $\gamma_{J}$ & $\gamma_{J} / \gamma_{M}$ \\
	\hline
	\hline
$N$  & $\omrot = 0.0$ & $\eta_* \simeq 1.0184$ & 0.509 & 0.0 & 0.357 & 0.363 & -- & -- \\	
\hline
$A$	& $\omrot = 0.05$	& $\eta_* \simeq 1.0192$ & 0.510 & 0.016 & 0.364 & 0.358 & 0.870 & 2.43 \\
\hline
$B$	& $\eta = 1.02$	 & $\omrot_* \simeq 0.06804$ & 0.511 & 0.022 & 0.356 & 0.360 & 0.870 & 2.42 \\
\hline
$C$	& $\omrot = 0.1$     & $\eta_* \simeq 1.0220$ & 0.512 & 0.032 & 0.357 & 0.360 & 0.873 & 2.43\\
\hline
$D$	& $\eta = 1.035$	& $\omrot_* \simeq 0.2147$  & 0.520 & 0.070 & 0.357 & 0.360 & 0.872 & 2.42\\
\hline
$E$	& $\eta = 1.0505$	& $\omrot_* \simeq 0.2997$  & 0.5296 & 0.100 & 0.359 & 0.364 & 0.876 & 2.41 \\
\hline
$F$	& $\omrot = 0.3$	& $\eta_* \simeq 1.0506$  & 0.5296 & 0.100 & 0.360 & 0.362 & 0.888 & 2.45 \\
\end{tabular}
\caption{Summary of parameters and results for six different rotating sequences $A$ through $F$.  For each sequence we list which parameter we fix, the critical value of the parameter that is being varied, the ADM mass $M_{\rm tot}$ and angular momentum $J_{\rm tot}$ of the critical initial data, as well as results for the critical exponent $\gamma_{\rho}$ for subcritical data and $\gamma_{M}$ and $\gamma_{J}$ for supercritical data.  We also include results for the non-rotating limit, marked $N$, which have been obtained with a different numerical grid setup (see \cite{BauM15}).}
\label{Table1}
\end{table*}

As we discussed in more detail in \cite{BauM15}, our results for $\gamma_{M}$, $\gamma_{J}$ and $\gamma_{\rho}$ depend somewhat on which numerical data are included in the fit.  Close to criticality, where the evolution develops increasingly small features, the numerical solution becomes increasingly affected by numerical error.  We have confirmed that we can extend our results closer to criticality 
by using a higher grid resolution.  Too far from criticality, on the other hand, the results show deviations from the scaling relations (\ref{mass_scaling}), (\ref{rho_scaling}) and (\ref{ang_mom_scaling}), which hold only in the immediate vicinity of criticality.  Accordingly, we estimate our results for the critical exponents to be accurate to within only a few percent.  

Within these error bars, our results for the critical exponents $\gamma_\rho$, $\gamma_{M}$ and $\gamma_{J}$ do not appear to be affected by the angular momentum of the initial data, which is consistent with the expectations from perturbative treatments \cite{Gun02,GunB16} as well as the numerical findings of \cite{ChoHLP04}.  Moreover, our findings for $\gamma_M$ and $\gamma_\rho$ are consistent with the analytical value of  $\gamma_M = \gamma_\rho \simeq 0.3558$ \cite{KoiHA95,Mai96}, while our results for $\gamma_J$ are close to the analytical value of $\gamma_J \simeq 0.8895$ \cite{Gun02}.   We also note that we obtain consistent values for these exponents independently of whether we vary $\eta$ or $\omrot$, i.e.~independently of the ``direction" in which the critical curve in Fig.~\ref{Fig:sketch} is crossed (see also \cite{GunB16}). 

\begin{figure}[t]
\includegraphics[width=3.5in]{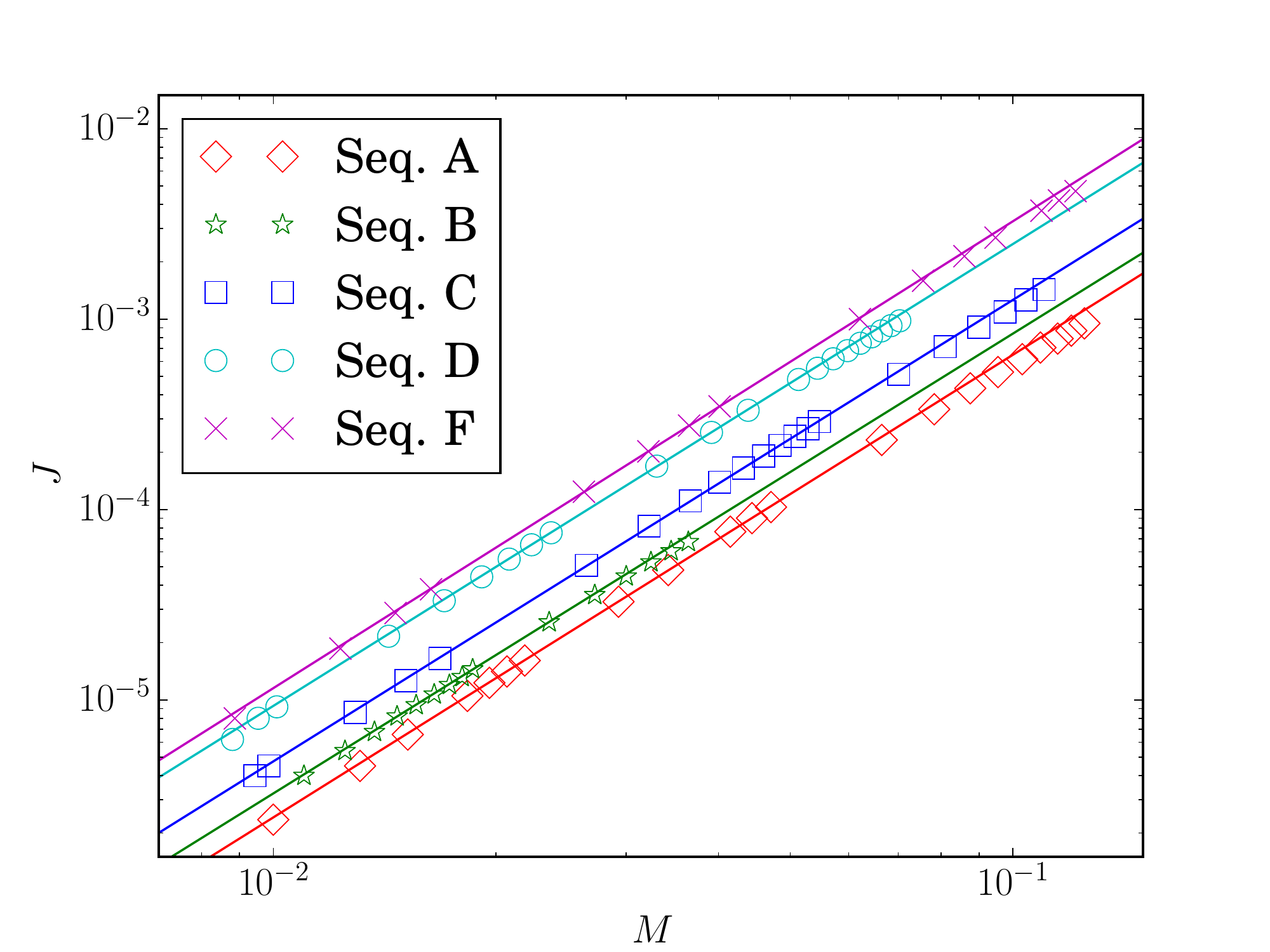}
\caption{Graphs of the angular momentum $J$ versus the mass mass $M$ of black holes formed in our supercritical evolutions.  The dots denote numerical results, while the solid lines are fits based on the parameters listed in Table~\ref{Table1}.  The slope of these lines is given by $\gamma_J/\gamma_M$ (last column in Table \ref{Table1}), which are close to the analytical value of 2.5 for a radiation fluid.}
\label{Fig:Fig3} 
\end{figure}

In Fig.~\ref{Fig:Fig3} we graph the angular momentum $J$ versus the mass $M$ of black holes formed in supercritical sequences.  As expected from (\ref{J_versus_M}) we find a power-law relation between these two quantities, with the exponent given by $\gamma_J/\gamma_M$.  Our numerical values for this ratio, listed in the last column in Table~\ref{Table1}, are close to the analytical value
$\gamma_J/\gamma_M = 2.5$ for a radiation fluid according to the perturbative treatment of \cite{Gun02} (see eq.~(\ref{gammas}) above).  

\begin{figure}[t]
\includegraphics[width=3.5in]{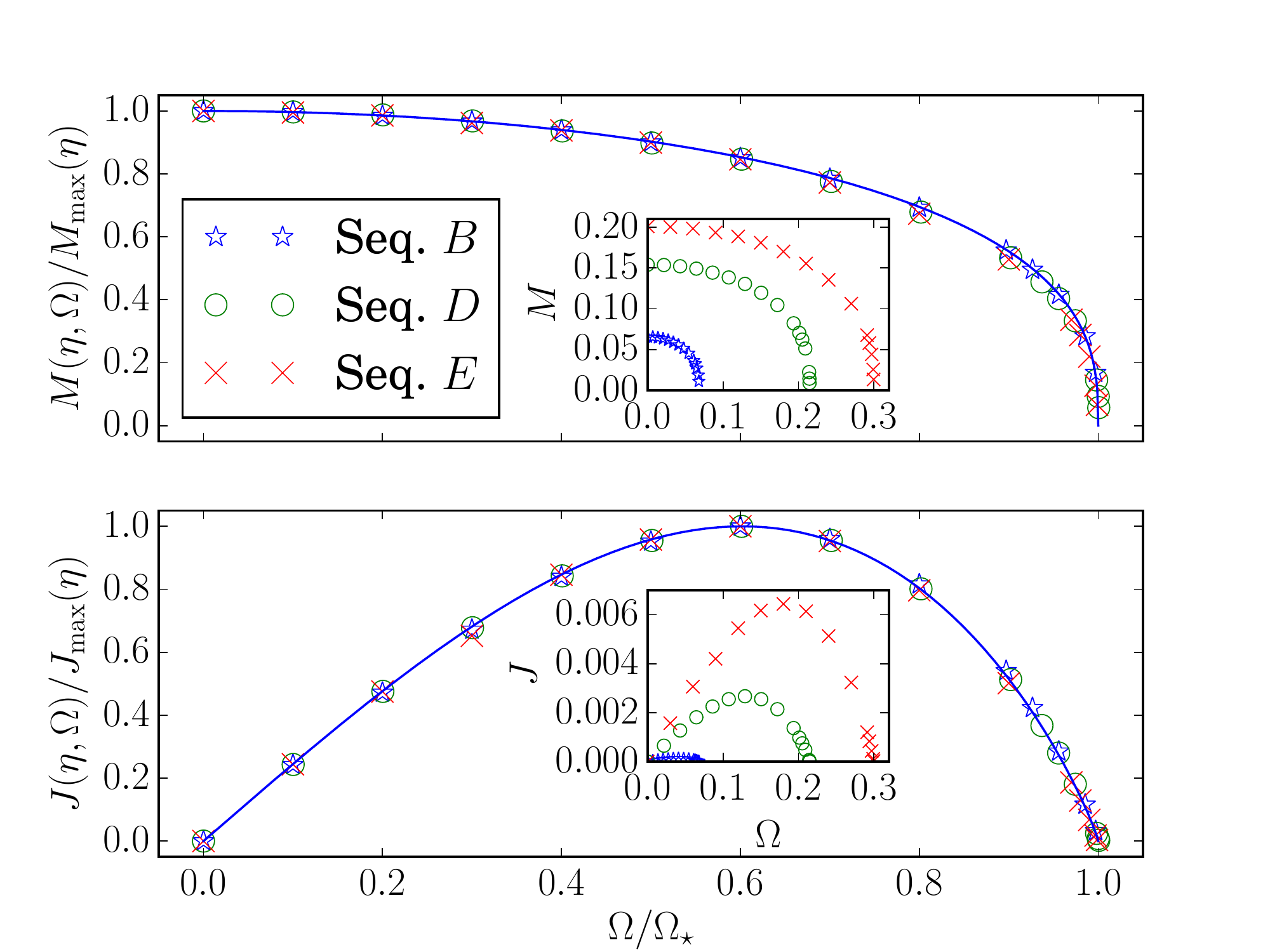}
\caption{Black hole masses $M$ and angular momenta $J$ as a function of $\eta$ and $\omrot$ for sequences $B$, $D$ and $E$.  The dots denote numerical results, while the solid lines are the analytical power-law scalings (\ref{scaling_function_mass}) and (\ref{scaling_function_ang_mom}).  The insets show the raw data for $M$ and $J$ as a function of $\omrot$, rather than the rescaled data as in the main graphs.}
\label{Fig:Fig4} 
\end{figure}

We now explore the supercritical ``horizontal" sequences $B$, $D$ and $E$ in Fig.~\ref{Fig:sketch} between $\omrot = 0$ and $\omrot_*$.   As shown in \cite{GunB16}, the simple power laws (\ref{mass_scaling}) and (\ref{ang_mom_scaling}) can be extended to provide closed-form expressions for the mass and angular-momentum of rotating black holes formed not only close to criticality, as predicted in \cite{Gun02,Gun98b}, but in the entire supercritical region shown in Fig.~\ref{Fig:sketch}.  Normalizing with respect to the maxima $M_{\rm max}(\eta)$ and $J_{\rm max}(\eta)$ of the mass and angular momentum for a given value of $\eta$ (with $M_{\rm max}(\eta) = M(\eta,0)$) we have
\begin{equation} \label{scaling_function_mass}
\frac{M(\eta,\omrot)}{M_{\rm max}(\eta)} \simeq (1 - x^2)^{\gamma_{M}}
\end{equation}
and 
\begin{equation} \label{scaling_function_ang_mom}
\frac{J(\eta,\omrot)}{J_{\rm max}(\eta)} \simeq \frac{x (1 - x^2)^{\gamma_J}}{C}.
\end{equation}
Here we have defined $x \equiv \omrot / \omrot_*(\eta)$ (where $\omrot_*(\eta)$ can be found by inverting (\ref{eta_crit_fit})), and $C \simeq 0.4025$ is the maximum of the function $x(1-x^2)^{\gamma_J}$ on the interval $[0,1]$.   Close to the critical curve, i.e.~in the limit $x \rightarrow \pm 1$, eqs.~(\ref{scaling_function_mass}) and (\ref{scaling_function_ang_mom}) reduce to (\ref{mass_scaling}) and (\ref{ang_mom_scaling}).


In Fig.~\ref{Fig:Fig4} we compare our numerical results with the expressions (\ref{scaling_function_mass}) and (\ref{scaling_function_ang_mom}).  Even though the masses and angular momenta themselves take vastly different values along the different sequences (see the insets in Fig.~\ref{Fig:Fig4}), they agree remarkably well when rescaled as suggested by (\ref{scaling_function_mass}) and (\ref{scaling_function_ang_mom}), especially for the sequences closer to $\eta_{*0}$.  

The maximum value of $J/M^2$ achieved on a line of constant $\eta$ scales with $(\eta - \eta_{*0})^{\gamma_M/2}$  \cite{GunB16}.  In the parameter region of Fig.~\ref{Fig:sketch}, the largest values of $J/M^2$ are about 0.29.  We therefore expect deviations from power-law scalings for larger values of $\eta$ in order to avoid violations of the constraint $J/M^2 <1$.  We plan to explore this regime in future work.

To summarize, we report on what we believe are the first fully relativistic simulations of the gravitational collapse of rotating radiation fluids.  We consider different sequences in our two-dimensional parameter space, and locate the critical curve separating supercritical from subcritical data.  We observe critical scaling of the black hole mass and angular momentum for supercritical data, and of the maximum encountered density for subcritical data.  The critical exponents are in good agreement with the perturbative results of \cite{Gun02}, are not affected by angular momentum, and are also universal in the sense that they do not depend on the direction in which the critical curve is crossed.   Our findings confirm that the angular momentum decreases more rapidly than the square of the black hole's mass as criticality is approached, so that in this limit the black hole is non-spinning.  We also demonstrate that, for supercritical data, the black hole masses and angular momenta satisfy new closed-form power-law scalings \cite{GunB16}.  Our findings therefore confirm several results on the role played by angular momentum in the critical gravitational collapse of radiation fluids, that, to date, had only been predicted from perturbative calculations.  We expect that our results do not depend on the specific choice of initial data, but that remains to be verified in future studies.

\acknowledgements

CG would like to thank Silvestre Aguilar-Martinez for helpful conversations.  This work was supported in part by NSF grant PHY-1402780 to Bowdoin College.

%

\end{document}